\documentclass[preprint,preprintnumbers,amsmath,amssymb]{revtex4}
\makeatother
\usepackage[dvips]{graphicx}
\usepackage{amsmath}
\usepackage{amsbsy}
\usepackage{caption}
\usepackage{color}
\DeclareGraphicsExtensions{.pdf,.eps,.png,.jpg,.mps}

\definecolor{textcolor}{cmyk}{0,0,0,1}
\definecolor{magenta}{rgb}{1,0,1}
\definecolor{green}{rgb}{0,1,0}
\definecolor{red}{rgb}{1,0,0}

\begin{document}

\title{
Substitutional 4d and 5d Impurities in Graphene
}
\author{ T. Alonso-Lanza,$^{1}$ A. Ayuela,$^{1}$ F. Aguilera-Granja,$^{1,2}$}
\affiliation{
$^1$Centro de F\'{\i}sica de Materiales CFM-MPC CSIC-UPV/EHU, Donostia 
International Physics Center (DIPC), Departamento de F\'{\i}sica de Materiales, Fac. de Qu\'{\i}micas, UPV-EHU, 20018 San Sebasti\'an, Spain
\\
$^2$Instituto de F\'{\i}sica, Universidad Aut\'onoma de San Luis de Potos\'{\i}, 78000 San Luis Potos\'{\i} S.L.P., M\'exico
\\
}

%\date{\today}

\begin{abstract}
We describe the structural and electronic properties of graphene doped with substitutional impurities of 4d and 5d transition metals. The binding energy and distances for 4d and 5d metals in graphene show similar trends for the later groups in the periodic table, which is also well-known characteristic of 3d elements. However, along earlier groups the 4d impurities in graphene show very similar binding energies, distances and magnetic moments to 5d ones,  which can be related to the influence of the 4d and 5d lanthanide contraction. Surprisingly, within the manganese group, the total magnetic moment of 3$\mu_{B}$ for manganese is reduced to 1$\mu_{B}$ for technetium and rhenium. We find that with compared with 3d elements, the larger size of the 4d and 5d elements causes a high degree hybridization with the neighbouring carbon atoms, reducing spin splitting in the d levels.
It seems that the magnetic adjustment of graphene could be significantly different is 4d or 5d impurities are used instead of 3d impurities.  
\end{abstract}

\maketitle

\section{Introduction}
%Gomes et al(REF ANDRES) present a simple model that let understand the bonding of the impurity atom with the carbon atoms.
%Based on this model, there are three different regimes, depending on the degree of filling of the 3d orbital.
%However, as for iron, the limit between different regimes it is not clear. The magnetic moment of this case is highly dependent with the U term or the iron carbon distance.

% grafeno
The isolation of a single layer of graphene in 2004\cite{novoselov2004electric} was followed by large body of works that sought to characterize this new material. Graphene stands out not only for its remarkable mechanical and optical characteristics but also for its electronic properties \cite{hua2013graphene,novoselov2012roadmap,huang2011graphene,geim2009graphene,neto2009electronic}. Charge carriers in graphene behave like Dirac fermions, which means that they can travel a few microns at room temperature without suffering the influence of scattering. It is this high mobility of electrons that makes graphene such a promising material for electronic applications.
The absence of a band gap, which is a drawback for its use in electronic applications, can be tuned using defects, for instance by chemical functionalization \cite{lawlor2014sublattice,liu2011chemical}. Furthermore, the magneto-conductivity signal can be detected by following the state of the magnetic molecules deposited over the graphene \cite{candini2011graphene}.
Its use in spintronics \cite{roche2015graphene} is thanks to the low spin-orbit interaction of carbon and the long spin coherence length of graphene\cite{tsukagoshi1999coherent}. The injection of the spin current in graphene has been proposed via ferromagnetic cobalt and nickel contacts  \cite{maassen2010graphene,lazic2014graphene}, and a perfect spin filter effect has been found for doped graphene nanoribbons\cite{rocha2010disorder}. 
The focus of the present work is on the substitutional doping of defective graphene with transition metals in order to study their effect on its magnetic properties.

% defectos intrínsecos
Although the non-magnetic behavior of pristine graphene appears to constitute a barrier to many applications, the creation of defects can be a method of providing a magnetic moment\cite{yazyev2010emergence,banhart2010structural}. Defects can furthermore be classified into intrinsic and extrinsic. Among the possible intrinsic defects\cite{dai2011first,lopez2009magnetic,ma2004magnetic,palacios2008vacancy,lusk2008nanoengineering} are vacancies, divacancies, dislocations, Stone-Wales defects, and domain defects. STM experiments have shown the relationship between the vacancies and magnetism as moderated by the presence of states close to the Fermi energy\cite{ugeda2010missing}, in agreement with theoretical calculations\cite{pereira2006disorder}. Control of the magnetism of a vacancy by strain has also been proposed\cite{santos2012magnetism}. Magnetic moments of vacancies have also been studied for graphene bilayers, where there is a reduction due to interlayer charge transfer\cite{choi2008monovacancy}.

When creating vacancies $\pi$ magnetism is developed, which could present a long-range magnetic order that could be ferromagnetic or antiferromagnetic depending on the defect sublattice. Even non-magnetic defects have been predicted to display a short-range magnetic order\cite{kumazaki2007nonmagnetic}.
Measurements of the magnetic states of graphene have produced contradictory results. A magnetic signal has been found for carbon systems such as graphite nodules\cite{coey2002ferromagnetism}, rhombohedral $C_{60}$\cite{makarova2001magnetic} and schwarzite-like structures\cite{park2003magnetism}, as well as a weak ferromagnetic signal at room temperature for graphene samples obtained from graphene oxide\cite{wang2008room}. However, a recent experiment\cite{sepioni2010limits} showed that there is no ferromagnetic interaction between defects, a finding supported by calculations\cite{palacios2012critical}.
This kind of magnetism arising from zigzag edges is also relevant for nanographenes, where local moments appear \cite{kumazaki2008local,sheng2010magnetism,fernandez2007magnetism,radovic2005chemical} when more than three zigzag units are found in the straight edges \cite{bhowmick2008edge}. It has been proposed that this edge-states magnetism could be reduced by various different mechanisms\cite{kunstmann2011stability}, but magnetic states have nevertheless been measured for zigzag nanographenes\cite{chen2013towards}.

% defectos extrínsecos
Extrinsic defects include adatoms and substitutional impurities, which can be placed both on vacancies and divacancies\cite{tsetseris2014substitutional}. 
Even carbon adatoms have been found to produce a localized magnetic moment\cite{lehtinen2003magnetic}.
Researchers have focused on the interaction of transition metal atoms interacting with both graphene\cite{faizabadi2013density,boukhvalov2011first} and graphene nanoribbons\cite{longo2011ab,power2011magnetization}. Osmium and iridium adatoms have been proposed to open a giant insulator topological gap in graphene\cite{hu2012giant}, while gold adatoms have been used to enhance spin-orbit splitting of $\pi$ bands\cite{ma2012strong}.
Interactions between graphene with both hydrogen atoms\cite{boukhvalov2008chemical,yazyev2007defect,lehtinen2004irradiation,forte2008modeling} and second period atoms\cite{sieranski2013energies,singh2009magnetism} have also been widely investigated. Moreover, experiments using 3d, 4d and 5d adatoms have been reported\cite{zolyomi2010first,zolyomi2010functionalization}.
The creation of extrinsic defect can be achieved by a two-step process: creation of the vacancy by irradiation, followed by the use of these defects as trapping centres for foreign atoms\cite{rodríguez2010trapping,wang2011doping}. There is large body of work describing the effects of irradiation of nanostructured materials\cite{krasheninnikov2010ion}. Raman spectroscopy has been revealed as a useful means of characterizing the presence of defects in graphene\cite{eckmann2013raman}. Nickel atoms trapped in graphene have also been studied using x-ray absorption fine structure (XAFS) analyses\cite{ushiro2006x}. % lo que hacemos nosotros
Previously, substitutional 3d transition metal atoms in a graphene layer were computed and a simple model was proposed to explain their bonding and magnetism\cite{krasheninnikov2009embedding,santos2010first}.  
Considering all these recent advances, it is clear that the missing research is therefore the study of the doping of 4d and 5d transition metals substitutionally in graphene.

Here we compute these substitutions in graphene in order to make comparison with the use of 3d atoms, and to check whether the previously proposed model can be generalized for all rows equally well.
We investigate the binding energies and geometries of the impurities and asses the magnetic behaviour of the doped graphene. Although it might be thought that 3d impurity results could be extrapolated to incorporate 4d and 5d transition metal atoms, there are some clear differences between them that justify a more detailed study.
It should be noted that only 3d atoms Fe, Co and Ni are ferromagnetics in bulk, according to the Stoner criterion. The wider bands of 4d and 5d atoms prevent them from having bulk ferromagnetic order at room temperature. However, at the nanoscale it is possible to find magnetic clusters for other species apart from these three traditional magnets \cite{kumar2003magnetism,moseler2001structure,reddy1999electronic,cox1993experimental,cox1994magnetism,liu1991magnetism}. As a result, the magnetic behaviour of these atoms in graphene needs careful consideration.
Furthermore, the fact that atom size increases when going following the periodic table down a group implies that size effects must play an important role when introducing impurities into graphene, because the atomic size determines the distances with the neighbouring carbons. Geometry appears the key to determining the magnetic state in some cases, revealing close relationship\cite{hughes2007lanthanide}. Where unexpected magnetic behaviour has been found, we have this explained in terms of a shell-adding phenomenon such as lanthanide contraction.
It seems that 4d and 5d impurities show less independence from graphene than 3d impurities, meaning that spin polarization may be better intertwined with the graphene bands.

\section{Computational details}

\begin{figure}[thpb]
      \centering
\includegraphics[width=8.5cm]{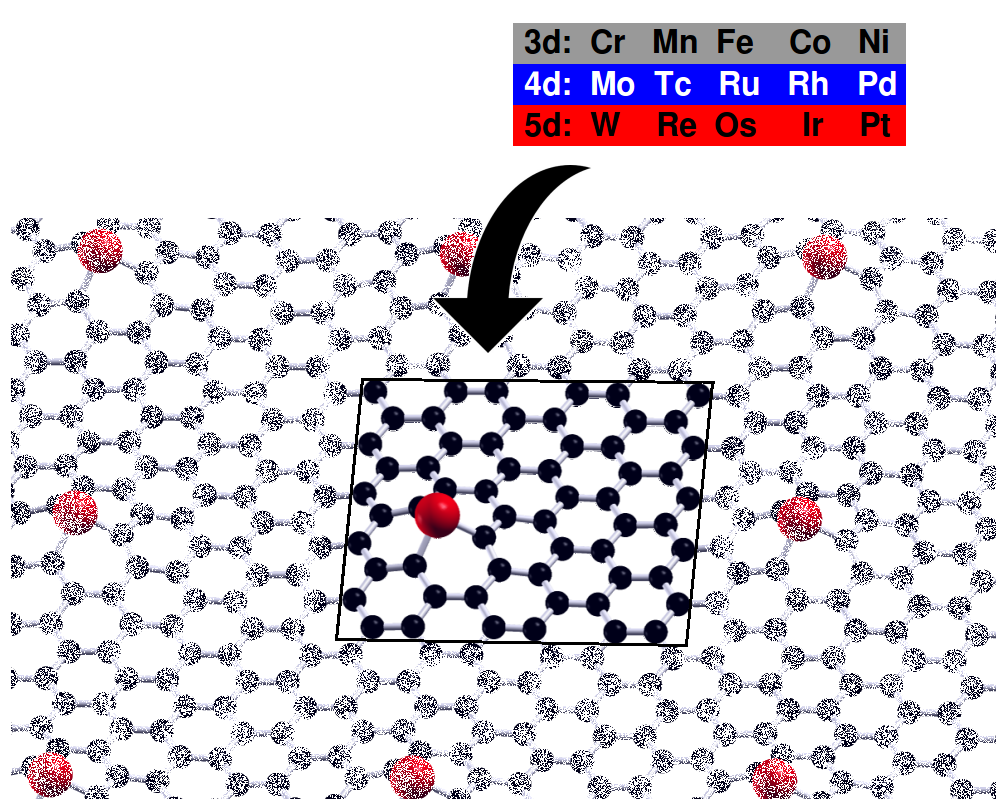}
\caption{\label{exp} 
Representation of the study described. Unit cell is highlighted by the rectangle.} 
   \end{figure}

% introducción y detalles generales
We carried out density functional theory (DFT) calculations on 4d and 5d substitutional impurities in graphene, using SIESTA (Spanish Initiative for Electronic Simulations with Thousands of Atoms) method.
For the exchange and correlation potentials we used the Perdew-Burke-Ernzenhof form of the generalized gradient approximation  (GGA)\cite{perdew1996generalized}.
The atomic cores were described by nonlocal norm-conserving relativistic Troullier-Martins pseudopotentials \cite{troullier1991efficient} with non-linear core corrections factorized in the Kleynman-Bylander form, the parameters of which are detailed in the appendix. 
% bases PONER TODO EN AMSTRONGS¿?
The basis size is double zeta plus polarization orbitals and the cutoff radii values are detailed in the appendix. 
%For the carbon atom, basis cutoff radii in bohrs are 5.949 and 3.519 in the 2s orbital and 7.638 and 3.889 in the 2p orbital.
Graphene is taken as a nearly square bidimensional unit cell of 12.9\AA{}$\times$12.5\AA{}, separated by an empty space between supercell replica converging to be non-interacting for a distance of 19 \AA{} as shown in Fig. \ref{exp}. We allowed the whole system composed of impurity plus graphene to relax, including the unit cell parameters, until the forces were less than 0.006 eV/\AA{}.
For all the calculations we use an electronic temperature of 25 meV and a meshcutoff of 250 Ry.
We used the Monkhorst-Pack method to choose the k-point mesh: 2x2x1, tested to converge with respect to the properties, as described below. 
We investigated pseudopotential transferability performing calculations for other kind of systems such as bulks and clusters, and using other codes.
We then also used the VASP (Vienna Ab initio Simulation Package) code in order to include the Hubbard term and to undertake GGA+U calculations for those cases where there can be some problem in the results obtained. We tried values of  $U_{eff} = U-J$ = 1, 2, 3 eV, representing a range of reasonable values as found in the literature\cite{piotrowski2011role,vaugier2012hubbard,wang2006oxidation,aykol2014local,jain2011formation} for transition metal systems.

\section{Results and discussion}

\subsection{
Binding energies and bond lengths}

\begin{figure}[thpb]
      \centering
\includegraphics[width=8.5cm]{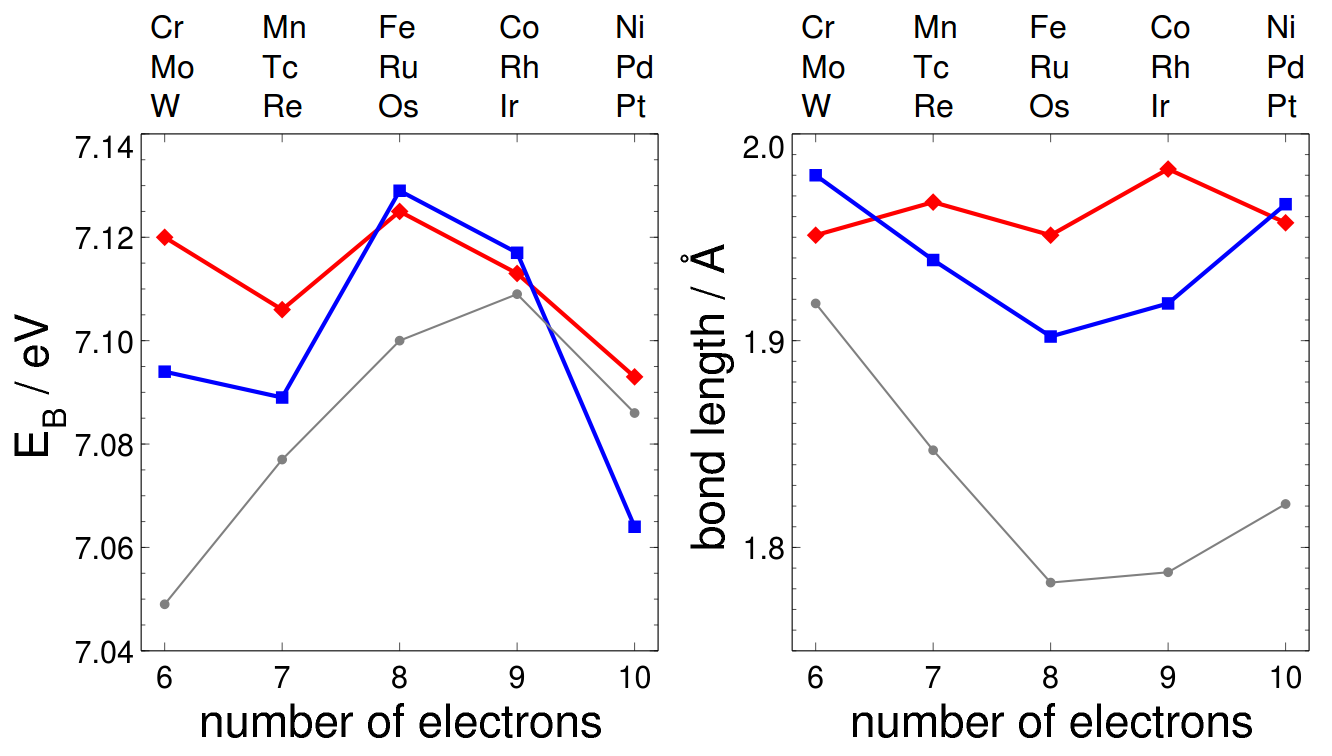}
\caption{\label{energy-distance} 
(a) Binding energies and (b) average bond lengths for the computed 3d (grey circles), 4d (blue squares) and 5d (red rhombuses) impurities in graphene.} 
   \end{figure}

In this section we describe the results for the binding energies and bond lengths.
First, we comment on the binding energies computed using the expression $E_{b}=n*E_{C}+E_{imp}-E_{total}$, taking into account the number of atoms $n$ in the unit cell and the energies of the different systems as per the subscript. The binding energies, shown in Fig. \ref{energy-distance} (a), are similar in all cases; all of these can be found within an interval of approximately 0.1 eV.
In general, the smallest values are found for 3d impurities except for palladium, which is less stable than nickel. 
Note that our 3d results for rectangular cells agree with previous values for a hexagonal cell distribution\cite{santos2010first}. The binding energies of the 4d impurities are similar to those of the 5d impurities. Moving from the manganese to iron group, the 4d impurities become slightly more stable; in the iron and cobalt groups, ruthenium and rhodium are more stable.
The variation in energy follows the next pattern in the 3d row: the binding energy grows from manganese to cobalt, and it next decreases for nickel. The largest stability is found for cobalt, in agreement with \cite{santos2010first}.
The variation in binding energy is different in the 4d row; technetium is less stable than would be expected if it were to follow the same trends as the 3d row. This also happens for rhenium in the 5d transition metals. Later, we associate this property with an abrupt change in the magnetic moment. It therefore seems that 4d and 5d elements differ from 3d elements where the d shells are half full.
Furthermore, the largest values of binding energy for 4d and 5d are found for ruthenium and osmium in the iron group, and not in the cobalt group as for 3d.
The most stable impurities in each row occur for cobalt in 3d, then in the iron group, in ruthenium (4d) and osmium (5d).

%%% FIGURE 2 %%%

% BOND LENGTH
For all the impurities the distances between the three nearest neighbour carbon atoms remain almost the same, with differences less than 0.004 \AA{}, so that the symmetry of the vacancy in graphene is well preserved. We calculated the average impurity-carbon atom bond length, as shown in Fig. \ref{energy-distance} (b).
The bond length for the 3d impurities decreases across the period as the atom size decreases, except for the last two magnetic elements, Co and Ni, for which the bond length shows a modest increase.
A similar trend is found for the bond lengths of the 4d impurities, but with larger values than for the 3d elements.
However, the bond lengths of the 5d impurities are similar around 1.97 \AA{} and show only small variations, within a range of 0.05 \AA{}.

The average bond length is closely related to the size of the impurity atom size. 3d impurities have a smaller atomic radius, which is correlated with their shorter bond lengths in graphene. Because 4d elements have a larger atomic radius than 3d elemnts, we expect greater bond lengths, in agreement with our results. The difference between the 3d and 4d case can be explained by the influence of the 4p electrons; in 4d elements these electrons screen the effective nuclear charge felt by the d electrons. Lower down in the periodic table, the difference in size between 4d and 5d elements is much less than between 3d and 5d elements. Note that for tungsten and platinum, the 4d and 5d order of the carbon-impurity distances within the group is actually reversed.
A similar behaviour affecting the 4d and 5d impurities is caused by the so-called lanthanide contraction\cite{taylor1972physics,seitz2007lanthanide,clavaguera2006calculated}. For 5d elements, the 4f levels have filled up before the addition of 5d electrons. After filling the 4f levels the electrons generally move far away from the nucleus, implying a poor shielding of the effective nuclear charge affecting the 5d electrons. The result is that 5d elements have a smaller radius than might be expected giving that a shell has been added, and the distances are similar to the 4d elements. Therefore, the lanthanide contraction can be indetified as the ultimate cause of the similar electronic behaviour between the 4d and 5d periods. This can also explain the previously mentioned trend in the carbon 5d-impurity bond lengths.

\subsection{
Magnetism}

%%% FIGURE 3 %%%
\begin{figure}[thpb]
\includegraphics[width=8.5cm]{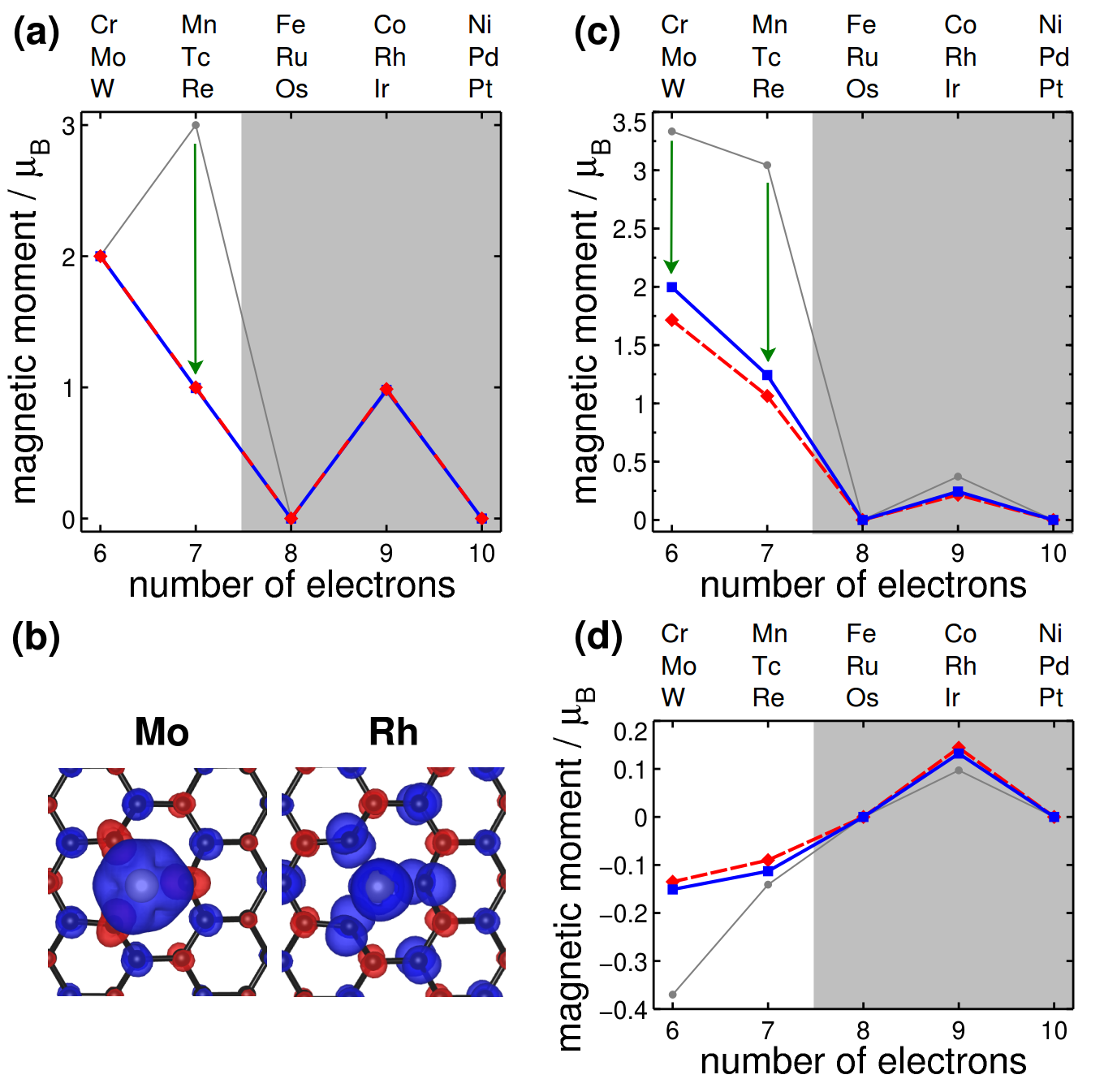}
\caption{\label{magnetism} 
Magnetic moments of graphene doped with 4d and 5d impurities, shown by blue squares and red rhombuses respectively. 3d impurities are indicated in grey circles for comparison.  (a) Total magnetic moments in the cell depicted using filled symbols joined with lines. (b) Spin densities for molybdenum and rhodium impurities, taken as examples of two magnetic regions. Positive and negative spin polarization are denoted in blue and red colours, respectively. (c) Local magnetic moments of impurity atoms. (d)  Induced graphene magnetic moments, taken as the averaged values for the three carbon neighbouring atoms. Note that for 4d and 5d impurities the magnetic moments are lower, as indicated by the down arrows in (a) and (b).  } 
\end{figure}

We now comment on the magnetism  of   4d and 5d elements in order to compare with the well-known data for the 3d cese.
The total magnetic moment is shown in Fig \ref{magnetism} (a) for all the transition metals considered here. In the case of graphene doped with 4d elements, the magnetic moment is 2$\mu_{B}$ for molybdenum, but 1$\mu_{B}$ for technetium.
The total magnetic moment in the 5d period follows that of the 4d shell, with values of 2, 1, 0, 1 and 0 $\mu_{B}$ for tungsten, rhenium, osmium, iridium and platinum, respectively. This shows a correlation between the number of electrons in the 4d and 5d shells and the values of magnetic moment, providing evidence of common properties within the same group.
Note that for 3d doping, the total cell magnetic moment increases from 2$\mu_{B}$ for chromium to 3$\mu_{B}$ for manganese, and alternates between the 0 and 1$\mu_{B}$ for the later transition metals nickel, cobalt, and iron \cite{santos2010first}, which is
at the border between both zones. Technetium (rhenium) matches the 0-1 alternating behaviour of the following Ru (Os), Rh (Ir) and Pd (Pt) elements, as in the case for the 3d elements in the later groups.

We distinguish two different regions in how impurities are aligned and graphene moments are induced, in terms of how full the orbitals are. Half-filled and less than half-filled elements in the chromium and manganese groups induce negative magnetic moments with respect to the three carbon neighbours, and in the cobalt group neighbouring carbon atoms have a positive induced magnetic moment, as already discovered for 3d elements \cite{santos2010first}. The spin density of molybdenum and rhodium impurities is plotted in Fig. \ref{magnetism}(b) to illustrate this for the case of  4d elements. 
Thus, magnetism for early d elements mainly depends on the impurity atom, but in the later d elements it is more delocalized over the graphene carbon atoms.
In all cases, the induced magnetism on graphene caused by to the impurity follows the sublattice AB pattern in graphene, with a negative magnetic moment on one sublattice and a positive moment on the other. The magnetism on the three carbon neighbours of the impurity define the nearest sublattice pattern; therefore, the sublattice a  positive induced magnetic moment for molybdenum has a negative induced magnetic moment for rhodium. This finding follows what the standard model in p-d coupling systems such as magnetic semiconductors \cite{raebiger2004intrinsic}. 

We now analyse in more detail how these differences accumulate quantitatively in the total magnetic moment considering the local magnetic moments on the impurity and nearest carbon atoms, as shown in Figs \ref{magnetism} (c,d), respectively.   
For chromium and manganese the 3d impurity atom contributes a large local magnetic moment, of 3.33 and 2.93$\mu_{B}$ respectively. The chromium value is even higher than the total value, because it is compensated by the large induced magnetic moment of -0.38$\mu_{B}$ on the neighbouring carbon atoms, as indicated by the down arrows in Fig. \ref{magnetism} (a,c).
For manganese, the three carbon atoms contribute less, -0.11$\mu_{B}$. The magnetic moments between the Cr and Mn impurities, and the carbon atoms have opposite signs. A different behaviour is found for the cobalt impurity, where the magnetization for the impurity (0.37$\mu_{B}$) is aligned with those of three carbon neighbours, with 0.1$\mu_{B}$ at each carbon atom.
However, the local magnetic moment on the 4d impurity drops to 2$\mu_{B}$ for molybdenum and 1.25$\mu_{B}$ for technetium, and again the induced graphene magnetic moments on the neighbouring carbon atoms are negative with values of -0.15$\mu_{B}$ and -0.11$\mu_B$, respectively. 
For rhodium, the magnetic moment is 0.25$\mu_{B}$, and the magnetic moment of the carbon atoms is positive, at 0.13 $\mu_{B}$ for each neighbouring carbon atom.
The local magnetic moments associated with the 5d impurity are even lower than the 4d case, at 1.72$\mu_{B}$ and 1.06$\mu_{B}$ for tungsten and rhenium respectively. The local value for iridium is 0.22$\mu_{B}$.
The induced response on the carbon atoms is similar to that for the other periods above: the carbon atoms have negative magnetic moment of -0.13$\mu_{B}$ for tungsten and -0.09$\mu_{B}$ for rhenium, and positive for the iridium impurity,  0.15$\mu_{B}$. Note that for the cobalt group the local magnetic moment in the impurity is reduced and the contribution from carbon neighbours increases down the group, yielding an increasing delocalization of the magnetic moment.
For the early d-groups, the drops in the local magnetic moments of both the impurities and the nearest carbon atoms  highlight the underlying differences  between the 3d and the 4d-5d rows.

\subsubsection{Manganese group magnetism}

%%% FIGURE 4 %%%
\begin{figure*}[thpb]
      \centering
\includegraphics[width=15cm]{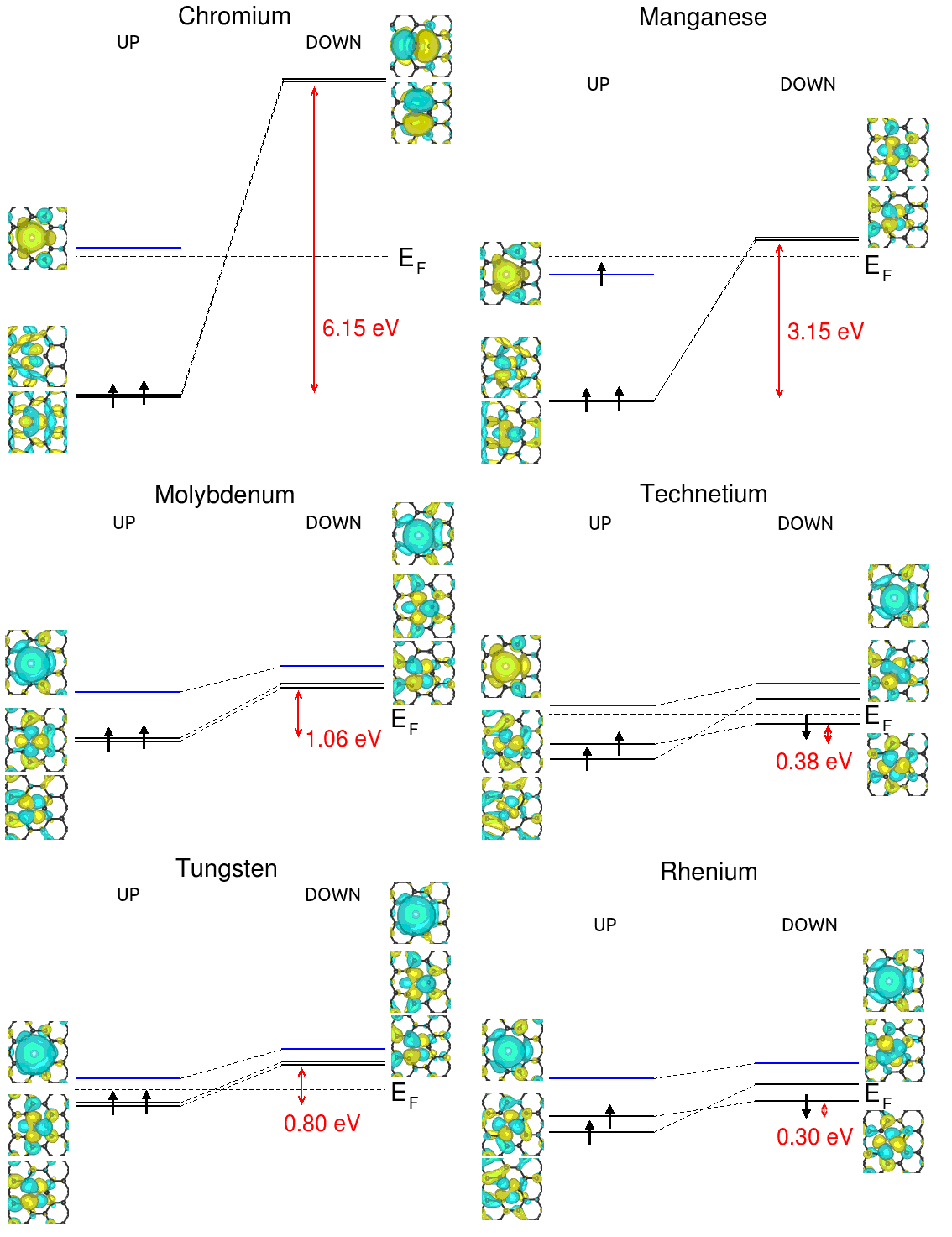}
\caption{\label{dos} 
Scheme showing levels close to the Fermi energy at the gamma point for (left) Cr, Mo, and W, and for (right) Mn, Tc, and Re impurities in graphene. The A (E) symmetry levels are indicated in blue (black). The respective wavefuntions are shown to the side of each level. The spin splitting values are marked on the energy scale by vertical red lines with double-headed arrows.} 
\end{figure*}

More importantly, in Fig \ref{magnetism} the values of the total magnetic moments of the cell show good agreement between 3d, 4d and 5d impurities, except for the manganese group. For manganese, the total magnetic moment is 3$\mu_{B}$, while it is 1$\mu_{B}$ for technetium and rhenium. 
%This is highlighted in Fig. \ref{magnetism} by the vertical jump.
Note that any possible difference between the 3d and 4d elements should be equally apply to 5d elements. 
%SPIN SPLITTING
To explain these differences especially in the manganese group, Fig. \ref{dos} shows a scheme of the levels close to the Fermi energy for the impurities of the chromium and manganese group. Since the symmetry $C_{3v}$ of the vacancy is well preserved, we consider the electronic levels classified into the A or E irreducible representation. An A symmetry level is a hybrid state formed by dz$^{2}$, s and p$_{z}$ states, and an E symmetry down level is assimilated to the rest of the d states of the impurity. 
In the chromium group, there are two possible options for the extra electron for the manganese group impurities: it can either fill an A symmetry up level or it can fill an E symmetry down level. 
For technetium the LUMO has an A symmetry as it can be seen in Fig. \ref{dos}.
Filling up the levels with A symmetry implies increasing the total magnetic moment to 3$\mu_{B}$, and this is what happens in manganese. However, the solution with low magnetic moment for technetium and rhenium fills the E symmetry down level, reducing the total magnetic moment to 1$\mu_{B}$.

The key point is that the spin splitting balances the energy needed to promote an electron to the A symmetry level. For manganese, the spin splitting is large enough to overcome the energy cost of promoting the electron to the A symmetry level. However, for technetium and rhenium, the spin splitting is small and the cost of promoting the electron to the upper level is decisive in avoiding the high-spin solution.
We also verify that the spin splitting in the chromium group is largely reduced for molybdenum and tungsten, as shown in the right-hand panels of Fig. \ref{dos}.

We now consider the detail of the hybridization of the d impurities in the composition at the level with A symmetry. 
For 3d elements, such as chromium and manganese, the largest contribution to that level comes from the p$_{z}$ states of the neighbouring carbon atoms; a smaller contribution comes from the impurity states. However, for the 4d elements  such as molybdenum and technetium, and for the 5d elements, tungsten and rhenium, the largest contribution comes from the d$_{z^{2}}$ states.
Again, we found that the impurity A-level hybridization is similar for 4d and 5d cases. This is related to the increasing size of the impurity when going down a group, which means in electronic terms that 4d and 5d electrons are more shielded by internal electrons than 3d electrons. Equally, because as the atoms of chromium group elements are larger than those in the manganese group, the contribution of the d$_{z^{2}}$  states is larger for the former.
The hybridization with carbon atoms is thus greater for 4d and 5d impurities. When the hybridization increases, not only is the atomic character of the impurity partially lost, but the spin splitting is also greatly reduced. 

In order to test this theroy, we computed by other method the technetium impurity with fixed spin moments of 3$\mu_{B}$ because it seems that this element lies on the border between regimes. Note that the equivalent case for 3d elements is iron \cite{santos2010first}. After relaxation, the bond lengths are slightly greater (0.03\AA{}), which implies less hybridization. This support the fact that the largest contribution to the now occupied A-symmetry level comes from the p$_{z}$ carbon states. 
Equally, if we set the impurity-carbon atom distance to be slightly larger (at 1.97\AA{} instead of 1.94\AA{}) and we do not relax, we obtain the result that the total magnetic moment is 3$\mu_{B}$. Note the delicate balance between distance and magnetism\cite{hughes2007lanthanide}. In any case, our results for technetium  in which the Coulomb term is strengthened for reasonable values of the U parameter shows that the total magnetic moment still remain with 1 $\mu_B$, as discussed above. 
Note that there is just a slight redistribution in local magnetic moments enforced by the inclusion of the U parameters.
The result is in good agreement with the fact that for ruthenium and osmium there is no change in the total magnetic moment even if the Coulomb term is included, which differs from the result where the 3d iron gains magnetization for the reasonable values of U \cite{santos2010first}. Those results are a proof of the considerable difference between 3d and 4d elements and of the similarity between 4d and 5d elements.

%\subsection{DFT+U}

%As it was seen in the case of iron \cite{santos2010first}, the U term could be of capital importance, modifying completely the magnetic moment of the system. For this reason, it seems logic to look at the border cases for the 4d and 5d elements, in order to shed more light over the problem.
%For strongly correlated systems it is necessary to include the on-site interaction to faithfully represent the real system. This is sometimes the case of the d electrons in transition metal atoms. In order to check the influence of on-site interaction on the systems under study here we carried out DFT+U calculations using the VASP code. We computed all the systems again using the relaxed geometries obtained with SIESTA including a U-J difference of 1, 2 and 3 eV.

\section{Summary and conclusions}

We found that 4d and 5d substitutional impurities in graphene behave similarly in terms of their geometric and magnetic properties, and notably differently from 3d impurities. This is supported by the binding energy and bond length trends presented here. We also found an abrupt difference in the magnetic moment of the cell between manganese and technetium.
	 Among the differences with respect to 3d impurities are that there are localized magnetic moments in the early-element 4d and 5d impurities with parallel moments to the three closest carbon atoms. They have more hybridized magnetic moments with both the impurity and the three carbon neighbours showing majority spin contributions.
There is a gerater hybridization with graphene for 4d elements compared to 3d elements, due to the reduction in the spin splitting. It appears that these 4d and 5d impurities could provide enhanced spin injection properties when doped substitutionaly in graphene.

\begin{acknowledgments}
This work has been partially supported by the Project FIS2013-48286-C2-1-P of the Spanish Ministry of Economy and Competitiveness MINECO, by the Basque Government under the ETORTEK Program 2014 (nanoGUNE2014), and the University of the Basque Country (Grant No. IT-366-07). TAL acknowledge a grant provided by the MPC Material Physics Center - San Sebasti\'an. FA-G would like to thank the DIPC for their generous hospitality. We also acknowledge the assistance provided by the DIPC computer center (TAL, AA and FA-G).

\end{acknowledgments}

\end{document}